\documentclass[10pt, conference, letterpaper]{IEEEtran}

\usepackage[noadjust]{cite}

\usepackage{subfigure}
\usepackage{amsmath}
\usepackage{amsthm}
\usepackage{graphicx} 
\usepackage{caption}
\usepackage{graphicx,subfigure}
\usepackage{multicol} 
\usepackage{graphicx} 
\usepackage{amssymb}
\usepackage{float} 
\usepackage{wrapfig} 
\usepackage{booktabs}
\usepackage{lipsum} 
\usepackage{indentfirst}
\usepackage{bm}
\usepackage{color}
\usepackage{amsmath}
\usepackage{enumerate}
\usepackage{extarrows}
\usepackage{url}

\usepackage{graphicx}
\usepackage{epstopdf}
\usepackage{multirow}
\usepackage{tablefootnote}
\usepackage[flushleft]{threeparttable}
\usepackage{slashbox}
\usepackage{array}
\usepackage{amsmath}
\usepackage{enumitem}
\usepackage{framed}

\usepackage{tcolorbox}

\usepackage[ruled,linesnumbered,vlined]{algorithm2e}  
\usepackage{framed}

\IEEEcompsocitemizethanks
\IEEEcompsocthanksitem
\IEEEoverridecommandlockouts

\usepackage{lscape}

\newtheorem{corollary}{\textbf{Corollary}}

\newtheorem{definition}{Definition}
\newtheorem{theorem}{\textbf{Theorem}}

\newtheorem{auction}{Auction}

\newtheorem{goal}{Requirement}
\newtheorem{baseline}{Baseline}

\usepackage{hyperref}
\hypersetup{hidelinks}

\definecolor{darkgreen}{rgb}{0.13, 0.55, 0.13}

\def\reward{r}
\def\cost{c}
\def\QFair{\lambda}
\def\thrFair{\phi}
\def\action{a}
\def\step{\eta}

\begin{document}

\title{Socially-Optimal Mechanism Design for Incentivized Online Learning}

\author{
	\IEEEauthorblockN{
		Zhiyuan Wang\IEEEauthorrefmark{1},
		Lin Gao\IEEEauthorrefmark{2}\IEEEauthorrefmark{4}, and
		Jianwei Huang\IEEEauthorrefmark{3}\IEEEauthorrefmark{4}
		\thanks{
			This work is supported by the National Key R\&D Program (NKRDP) of China under Grant 2019YFB1802803.
			This work is supported by the Shenzhen Science and Technology Program (JCYJ20210324120011032), Shenzhen Institute of Artificial Intelligence and Robotics for Society, and the Presidential Fund from the Chinese University of Hong Kong, Shenzhen.
			This work is also supported by the National Natural Science Foundation of China (Grant No. 61972113), Shenzhen Science and Technology Program (Grant No. JCYJ20190806112215116, JCYJ20180306171800589, and KQTD20190929172545139), and Guangdong Science and Technology Planning Project under Grant 2018B030322004. (\textit{Corresponding author: Jianwei Huang.})}
	}
	
	\IEEEauthorblockA{\IEEEauthorrefmark{1}School of Computer Science and Engineering, Beihang University, Beijing, China}
	\IEEEauthorblockA{\IEEEauthorrefmark{2}School of Electronics and Information Engineering, Harbin Institute of Technology, Shenzhen, China}
	\IEEEauthorblockA{\IEEEauthorrefmark{3}School of Science and Engineering, The Chinese University of Hong Kong, Shenzhen}
	\IEEEauthorblockA{\IEEEauthorrefmark{4}Shenzhen Institute of Artificial Intelligence and Robotics for Society, Shenzhen, China}
}

\maketitle

\IEEEcompsocitemizethanks
\IEEEcompsocthanksitem
\IEEEoverridecommandlockouts

\begin{abstract}
	Multi-arm bandit (MAB) is a classic online learning framework that studies the sequential decision-making in an uncertain environment.
	The MAB framework, however, overlooks the scenario where the decision-maker cannot take actions (e.g., pulling arms) directly.
	It is a practically important scenario in many applications such as spectrum sharing, crowdsensing, and edge computing.	
	In these applications, the decision-maker would incentivize other selfish agents to carry out desired actions (i.e., pulling arms on the decision-maker's behalf).
	This paper establishes the incentivized online learning (IOL) framework for this scenario.
	The key challenge to design the IOL framework lies in the tight coupling of the unknown environment learning and asymmetric information revelation. 
	To address this, we construct a special Lagrangian function based on which we propose a socially-optimal mechanism for the IOL framework.
	Our mechanism satisfies various desirable properties such as agent fairness, incentive compatibility, and voluntary participation.
	It achieves the same asymptotic performance as the state-of-art benchmark that requires extra information. 
	Our analysis also unveils the power of crowd in the IOL framework: a larger agent crowd enables our mechanism to approach more closely the theoretical upper bound of social performance.
	Numerical results demonstrate the advantages of our mechanism in large-scale edge computing.
\end{abstract}

\section{Introduction}
\subsection{Background and Motivation}
\label{Subsection: Background and Motivation}
\begin{figure}
	\setlength{\abovecaptionskip}{0pt}
	\setlength{\belowcaptionskip}{0pt}
	\centering
	\subfigure[Online learning framework]
	{\label{fig: learning_1}\includegraphics[height=0.2\linewidth]{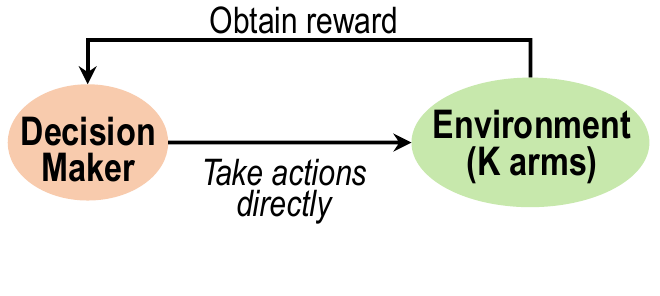}}\quad 
	\subfigure[IOL framework]
	{\label{fig: learning_2}\includegraphics[height=0.2\linewidth]{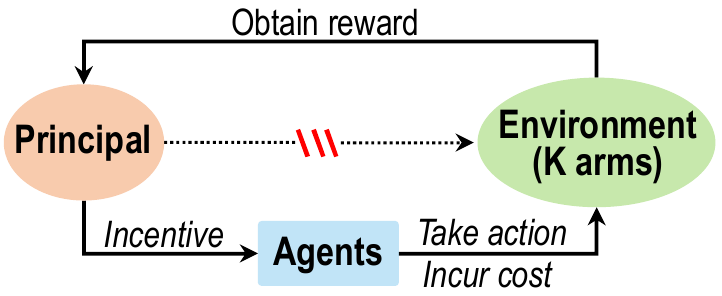}}
	\caption{Illustration of two frameworks}
	\label{fig: }
\end{figure}

Online learning has been widely adopted in many practical optimization problems, where the decision-maker takes sequential actions in an uncertain environment \cite{shalev2011online}.
Multi-arm bandit (MAB) is one of the extensively studied online learning frameworks \cite{bubeck2012regret}.
As shown in Fig. \ref{fig: learning_1}, the uncertainty in MAB corresponds to the unknown rewards that can be obtained from a set of arms.
The decision-maker gains reward by pulling an arm in each time slot, based on its observations of the realized reward from the arms pulled in previous time slots.
The goal is to maximize the cumulative reward over a time horizon.
This framework has a wide range of networking applications, such as wireless channel access (e.g., \cite{gai2012combinatorial,yang2015online,liu2015online}), crowdsourcing worker selection (e.g., \cite{Chao2021Eliciting,han2015taming,ul2016efficient,nee2018context,Wang2018Dynamic,Chao2021Stratigic}), and computation task offloading (e.g., \cite{chen2021distributed,ouyang2019adaptive,wu2020mab,Li2020Learning}).



Despite the wide adoption, the online learning framework misses an important scenario where the decision-maker is \text{\textbf{unable to take actions directly}}.
This is usually due to the fact that the decision-maker lacks the resource required to take actions.
In this scenario, the decision-maker would act as a principle and incentivize some selfish agents (who have resource) to take actions on his behalf.
The incentives are necessary, since the agents may incur costs when taking the actions.
In this paper, we will refer to this paradigm as the \textbf{incentivized online learning (IOL)} as shown in Fig. \ref{fig: learning_2}.
Such a principal-agent interaction widely exists in many real-world applications, such as spectrum sharing, mobile crowd sensing, and edge computing.
We briefly introduce two motivating examples shown in Table \ref{Table: examples}, and we will provide detailed demonstration in Section \ref{Section: Experiment Results}.

\begin{table}
		\setlength{\abovecaptionskip}{0pt}
		\setlength{\belowcaptionskip}{0pt}
	\renewcommand{\arraystretch}{1.4}		
	\caption{Motivating examples of IOL framework} 
	\centering
	\label{Table: examples}
	\begin{tabular}{|c|c|c|c|c|}
		\hline
		\textbf{Principal} 	& \textbf{Environment} (\textit{K} Arms)	& \textbf{Agents}	\\
		\hline\hline		
		IoT Operator	& $\normalsize\substack{\text{Sensors}\\\textit{(Unknown sensing data)}}$	& Spectrum owners \\
		\hline
		MCS Platform	& $\normalsize\substack{\text{Locations}\\\textit{(Unknown real-time  status)}}$	&	Mobile users \\
		\hline
	\end{tabular}
\end{table}

\begin{itemize}
	\item \textit{\textbf{Spectrum Sharing:}}
	The IoT operator aims to aggregate sensing data from multiple sensors, but may not have enough spectrum for the sensors to operate in \cite{zhang2018spectrum}.
	The IoT operator could incentivize other spectrum owners to share their spectrum with the sensors.
	The IoT operator's obtained reward depends on the sensing data amount and quality, which are unknown ex ante. 
	The spectrum owners perceive cost or disturbance on their own spectrum utilizations when sharing the spectrum with sensors, and such costs are spectrum owners' private information. 
	

	\item \textit{\textbf{Mobile Crowd Sensing (MCS):}}
	The MCS platform relies on mobile users to acquire the time-varying status of different locations \cite{ganti2011mobile}.
	The acquired status exhibits different values for the MCS platform, depending on the status changing and the importance of locations.
	Mobile users incur private cost when sensing the status. 
\end{itemize}
Based on the above discussion, let us move on to the IOL framework optimization.

\subsection{IOL Framework Optimization}
The online learning framework in Fig. \ref{fig: learning_1} aims to design an algorithm that maximizes the reward.
The IOL framework, however, is much more complicated due to the principal-agent interaction in Fig. \ref{fig: learning_2}.
Roughly speaking, the IOL framework requires us to design a \textit{mechanism} that properly manipulates the principal-agent interactions, thereby achieving various desirable prosperities (beyond reward maximization).
Next we introduce the properties in our mechanism design for the IOL framework (or simply IOL mechanism design).

First, the IOL framework in Fig. \ref{fig: learning_2} exhibits a novel learning paradigm that leverages the selfish agents.
The efficiency of this paradigm depends jointly on the principal's reward and agents' costs.
This motivates us to consider the following two properties in our mechanism design for the IOL framework:
\begin{definition}[Efficiency]\label{Property: Efficiency}
	A mechanism is efficient if it maximizes the social welfare in the IOL framework.
\end{definition}
\begin{definition}[Fairness]\label{Property: Fairness}
	A mechanism is fair if it ensures each agent a predefined maximal utilization in the IOL framework.
\end{definition}

Second, there is an incentive issue in the IOL framework.
On the one hand, the agents will not voluntarily take actions for the principal due to their costs.
On the other hand, the agents' costs are their private information unknown by the principle or other agents.
Therefore, our mechanism should help the principal tackle the information asymmetry and induce the agents to carry out the desired actions.
This leads to the following two properties in the IOL mechanism design.
\begin{definition}[Incentive Compatibility]\label{Property: IC}
	A mechanism satisfies incentive compatibility if it induces the agents to truthfully reveal their private costs in the IOL framework.
\end{definition}

\begin{definition}[Voluntary Participation]\label{Property: VR}
	A mechanism satisfies voluntary participation if it ensures both agents and principal will not be worse off by participating the IOL framework.
\end{definition}

The IOL framework involves coupled unknown environment and information asymmetry, the coexistence of which renders the corresponding mechanism design highly challenging.
To our knowledge, there is no systematic study yet.
In this paper, we will try to fill in the gap and propose a mechanism satisfying the properties in Definitions \ref{Property: Efficiency}$\sim$\ref{Property: VR}.
We hope that our study can inspire further research on the IOL framework.

\subsection{Main Results and Key Contributions}
We study the mechanism design problem for the incentivized online learning (IOL) framework.
Specifically, we characterize the principal-agent interactions in the uncertain environment based on stochastic MAB, and propose a socially-optimal mechanism satisfying several desired properties. 
The main results and key contributions are as follows:
\begin{itemize}
	\item \textit{A Unified IOL Framework:}	
	We propose an IOL framework motivated by multiple real-world applications, where the principal incentivizes selfish agents to perform the learning actions.
	This framework enriches the modeling flexibility of the classic online learning framework, and starts an unexplored learning paradigm that leverages the selfish agents.
	To the best of our knowledge, this is the first systematic study on this paradigm.
	
	\item \textit{Mechanism Design:}
	The major challenge in the IOL framework optimization is the coupled unknown environment and information asymmetry.
	To address it, we construct a special Lagrangian function based on the environment estimation and the information revelation.
	It enables us to properly integrate the learning and incentive rationale, and develop a mechanism satisfying several desired properties.
	Our proposed mechanism is incentive compatible, and ensures the voluntary participation.
	It asymptotically performs as well as the state-of-art baseline (that requires extra information) in terms of the social welfare and the agent utilization fairness.
	
	\item \textit{Power of Crowd:}
	We unveil the impact of the agent crowd size in the IOL framework.
	Our analysis shows that a larger agent crowd enables our proposed mechanism to approach more closely the theoretical upper bound of the social welfare.
	
	\item \textit{Numerical Results:}
	We demonstrate how to apply our proposed mechanism to the edge computing system.
	Results show that a larger edge network makes it less costly to incentivize the individual.
\end{itemize}

The rest of this paper is as follows.
Section \ref{Section: Related works} reviews related literature.
Section \ref{Section: Model} introduces the system model.
We propose an IOL mechanism and present the theoretical performance in Section \ref{Section: mechanism design} and Section \ref{Section: performance}, respectively.
Section \ref{Section: Experiment Results} provides the numerical results.
Section \ref{Section: Conclusion} concludes this paper.

\section{Related Works}\label{Section: Related works}
We will review two streams of literatures that are related to this paper in Section \ref{Subsection: Interplay} and Section \ref{Subsection: Generality}, respectively.

\subsection{Interplay of Exploration, Exploitation, and Incentive}
\label{Subsection: Interplay}
The IOL framework involves three-way interplay of exploration, exploitation, and incentive.
Previous studies on a series of problems known as \textit{incentivizing exploration} are also related to this interplay.
Next we highlight the key differences.

\subsubsection{Incentivizing Exploration}
The studies on incentivizing exploration are motivated by typical applications such as recommendation platforms (e.g., Yelp for restaurants) \cite{kleinberg2017tutorial}.
In these applications, the agent (e.g., a customer) gains reward by pulling an arm (e.g., eating in a restaurant), thus they will voluntarily take the exploitation action (e.g., going to a high-rating restaurant).
However, the principal (e.g., Yelp) is forward-looking, thus it will induce agents to adopt exploration (e.g., checking a low-rating restaurant) by providing incentive or compensation.
Frazier \textit{et al.} in \cite{frazier2014incentivizing} and Kremer \textit{et al.} in \cite{kremer2014implementing} initialized the basic formulation for incentivizing exploration based on the Bayesian MAB model and strategic information revelation, respectively.
Wang and Huang in \cite{wang2018multi} considered
the non-Bayesian MAB setting and derive the compensation lower bound.
Zhou \textit{et al.} in \cite{zhou2021incentivized} took into account the self-reinforcing user preferences, which capture the random user behaviors in most online recommender systems.
They also proposed two policies achieving $O(\log T)$ expected regret and $O(\log T)$ expected payment.

\subsubsection{IOL Framework}
Different from incentivizing exploration, the IOL framework focuses on the setting where the principal gains reward by persuading the agents to pull the arms.
In this setting, the agents will \textit{not voluntarily take the exploration-exploitation actions} for the principal due to their incurred costs.
Overall, the underlying three-way interplay in the IOL framework is substantially different from that in incentivizing exploration.

\subsection{Generality of IOL Framework}
\label{Subsection: Generality}
The IOL framework is widely applicable to many real-world applications.
Our proposed mechanism takes into account the coupled unknown environment and information asymmetry, as well as achieving several desired properties.
Our study generalizes several groups of prior works.
For example, Zhang \textit{et al.} in \cite{Zhang2018Caching} proposed an efficient delay-optimal cooperative mobile edge caching scheme, which ensures a provable optimality and low complexity.
The authors in \cite{Zhang2018Air} proposed a novel service-oriented network slicing approach to efficiently manage the multi-dimensional resource in air-ground integrated vehicular network.
Our mechanism can help tackle the unknown environment and information asymmetry in these resource sharing studies.
Furthermore, some studies on channel access (e.g., \cite{gai2012combinatorial,yang2015online,liu2015online}), crowdsensing (e.g., \cite{han2015taming,ul2016efficient,nee2018context,Wang2018Dynamic}), and edge computing (e.g., \cite{chen2021distributed,ouyang2019adaptive,wu2020mab,Li2020Learning}) mainly focused on unknown parameter learning.
For example, Wang \textit{et al.} in \cite{Wang2018Dynamic} proposed an efficient online control policy to address the stochastic characteristics and the discontinuous coverage in crowdsesing.
Li \textit{et al.} in \cite{Li2020Learning} designed a novel learning algorithm for cooperative edge computing, which ensures the close-to-optimal performance.
Our mechanism tackles the incentive issue from the selfish spectrum owners, crowdsensing workers, and edge severs, respectively.
Some studies (e.g., \cite{yang2016selecting,gao2021budgeted,xiao2021cmab}) also took into account the budget-constrained crowdsensing requester.
Our mechanism ensures voluntary participation for the principal, thus captures the same requirement.
Overall, we believe that our study in this paper is a promising attempt for establishing a systematic IOL framework.



\section{System Model}\label{Section: Model}
We study the mechanism design problem for the IOL framework shown in Fig. \ref{fig: learning_2}.
We first present the characterization for this framework in Section \ref{Subsection: IOL Framework Overview}.
We then introduce the corresponding mechanism design formulation in Sections \ref{Subsection: Participation}, \ref{Subsection: Fairness}, and \ref{Subsection: Efficiency}, respectively.

\subsection{IOL Framework Overview}
\label{Subsection: IOL Framework Overview}
We will characterize the IOL framework from two aspects:  the unknown environment (based on stochastic MAB) and the principal-agent interactions.

\subsubsection{Unknown Environment}
The classic stochastic MAB model defines the unknown environment based on the stochastic reward of a set $\mathcal{K}=\{1,2,...,K\}$ of arms \cite{bubeck2012regret}.
Each arm $k\in\mathcal{K}$ is associated with an independent and identically distributed (IID) reward.
We let $\reward_{k}^{t}\in[0,1]$ denote the reward of arm $k$ in slot $t$, and let $\bar{\reward}_{k}=\mathbb{E}[\reward_{k}^{t}]$ denote the corresponding average reward.
The principal will get the reward $\reward_{k}^{t}$ if arm $k$ is pulled in slot $t$.
We have two-fold elaboration on the principal:
\begin{itemize}
	\item First, the principal does not know the average reward vector $\bm{\bar{\reward}}=(\bar{\reward}_{k}:\forall k\in\mathcal{K})$, and it cannot observe the reward $\reward_{k}^{t}$ unless arm $k$ is pulled in slot $t$.
	
	\item Second, in some scenarios (e.g., Table \ref{Table: examples}), the principal cannot play the arms in person, which naturally leads to the principal-agent interactions in the IOL framework.
\end{itemize}

\subsubsection{Principal-Agent Interactions}
The IOL framework exhibits the strategic principal-agent interaction, where the principal will incentivize a set $\mathcal{N}=\{1,2,...,N\}$ of selfish agents to pull the arms for it.
The agents are self-interested and may incur costs from pulling arms for the principal.
The cost is the agent's private information and possibly time-varying.
Mathematically, we model the cost for each pair of agent $n$ and arm $k$ as the IID random variable $\cost_{n,k}^{t}\in[\cost_{\min},1]$, where $\cost_{\min}\ge0$ represents the potential minimal cost of taking an arm-pulling action.
Here we have two-fold elaboration on the principal-agent interaction:
\begin{itemize}
	\item First, the agent $n$'s cost realization $\bm{\cost}_{n}^{t}=(\cost_{n,k}^{t}:\forall k\in\mathcal{K})$ and the distribution are his private information, which are hidden from the principal and other agents.
	\item Second, agents are selfish, thus will not voluntarily disclose private information or pull arms for the principal.	
\end{itemize}

\subsubsection{Mechanism Design for IOL Framework}
Our goal in the IOL framework is to develop a mechanism that can help the principal incentivize the selfish agents to take the desired arm-pulling actions.
In each time slot $t$, our mechanism will generate a proposal $(\bm{x}^{t},\bm{y}^{t})$, which consists of the assignment $\bm{x}^{t}$ and the payment $\bm{y}^{t}$:
\begin{itemize}
	\item We let $\bm{x}^{t}=(x_{n,k}^{t}\in\{0,1\}:\forall n\in\mathcal{N},k\in\mathcal{K})$ denote the assignment matrix in slot $t$, where $x_{n,k}^{t}=1$ represents that agent $n$ is expected to pull arm $k$ for the principal.
	
	\item We let $\bm{y}^{t}=(y_{n}^{t}\ge0:\forall n\in\mathcal{N})$ denote the payment vector in slot $t$, where $y_{n}^{t}$ is the principal's monetary payment to agent $n$.
	To persuade agents to follow the assignment $\bm{x}^{t}$, it is critical to appropriately design $\bm{y}^{t}$.
\end{itemize}

Without loss of generality, we suppose that each agent can only pull one arm per slot and each arm can be pulled at most once per slot.
Mathematically, the assignment $\bm{x}^{t}$ is chosen from the set $\mathcal{X}$ defined as follows:
\begin{equation}\label{Equ: ActionSet}
	\begin{aligned}
		\mathcal{X}
		\triangleq
		\Big(
		\bm{x}\in\{0,1\}^{N\times K}\Big|\ &\textstyle \sum_{k=1}^{K}x_{n,k}\le1,\ \forall n\in\mathcal{N},\\
		&\quad\textstyle \sum_{n=1}^{N}x_{n,k}\le1,\ \forall k\in\mathcal{K}
		\Big).
	\end{aligned}
\end{equation}

Next we introduce four requirements on the mechanism design in Section \ref{Subsection: Participation}, \ref{Subsection: Fairness}, and \ref{Subsection: Efficiency}, respectively.
These requirements guide us to find the desired proposal $(\bm{x}^{t},\bm{y}^{t})$.

\subsection{Participation Requirement}\label{Subsection: Participation}
We start with the participation requirements on the agents and the principal.


\subsubsection{Agent Participation}
The agents are selfish and will decide whether to follow the proposal $(\bm{x}^{t},\bm{y}^{t})$ based on the payment and their incurred costs. 
Mathematically, we let $\bm{\action}^{t}=(\action_{n}^{t}\in\{0,1\}:\forall n\in\mathcal{N})$ denote the decisions of agents in slot $t$, where $\action_{n}^{t}=1$ and $\action_{n}^{t}=0$ represent that agent $n$ will follow and decline the proposal in slot $t$, respectively.
Given the proposal $(\bm{x}^{t},\bm{y}^{t})$, the payoff of agent $n$ in slot $t$ is 
\begin{equation}\label{Equ: Payoff}
	\begin{aligned}
		U_{n}(\action_{n}^{t},\bm{x}^{t},\bm{y}^{t})
		\triangleq  
		\left(y_{n}^{t}
		-
		\sum_{k=1}^{K}\cost_{n,k}^{t}x_{n,k}^{t}
		\right)\action_{n}^{t},
	\end{aligned}
\end{equation}
where $\sum_{k=1}^{K}\cost_{n,k}^{t}x_{n,k}^{t}$ represents the cost perceived by agent $n$ when he follows the assignment $\bm{x}^{t}$ in slot $t$.
To implement the assignment $\bm{x}^{t}$, the IOL mechanism should ensure that the payment $y_{n}^{t}$ to each agent $n\in\mathcal{N}$ is no less than the corresponding cost $\sum_{k=1}^{K}\cost_{n,k}^{t}x_{n,k}^{t}$.
In that case, the agents' payoff-maximizing decision is $\bm{\action}^{t}=\bm{1}_{N}$, where $\bm{1}_{N}$ is the $N$-dimensional all-one vector.
This leads to the first requirement in our mechanism design for the IOL framework.
\begin{goal}\label{Goal: agent_follow}
	To ensure the voluntary participation of all agents, i.e., $\bm{\action}^{t}=\bm{1}_{N}$, the mechanism should satisfy
	\begin{equation}\label{Equ: condition_follow}
	\textstyle
		y_{n}^{t}
		-
		\sum\limits_{k=1}^{K}\cost_{n,k}^{t}x_{n,k}^{t}
		\ge0,\quad\forall n\in\mathcal{N},t\in\mathcal{T}.
	\end{equation}
\end{goal}

\subsubsection{Principal Participation}
In the IOL framework, the principal pays the agents in exchange for the reward of arms played by the agents.
Therefore, the principal's profit in slot $t$ is the difference between the total reward and the total payment.
Given the assignment $\bm{x}^{t}$ and the agent decision $\bm{a}^{t}$, the principal gains the following reward in slot $t$:
\begin{equation}\label{Equ: valuation}
\begin{aligned}
V(\bm{x}^{t},\bm{\action}^{t};\bm{\reward}^{t})
\triangleq
\sum\limits_{k=1}^{K}\sum\limits_{n=1}^{N}
\reward_{k}^{t}x_{n,k}^{t}\action_{n}^{t},
\end{aligned}	
\end{equation}
where $\reward_{k}^{t}$ is the reward of playing arm $k$ in slot $t$.
Moreover, the principal's total payment is $\sum_{n=1}^{N}\action_{n}^{t}y_{n}^{t}$.

Intuitively, it is not beneficial for the principal to adopt the IOL framework if the reward is much smaller than the agents' costs.
However, the principal does not know whether this is the case until it gradually observes the reward realizations and elicits the agents' costs.
That is, it is inevitable that the principal may get an instantaneous \textit{deficit}, i.e., negative profit in some slot.
Therefore, we will focus on the principal's long-term profit in the IOL mechanism design.
For presentation convenience, we define the principal's \textit{cumulative profit} as
\begin{equation}\label{Equ: condition_profit}
	\begin{aligned}
		\mathtt{Pro}(T)
		\triangleq
		\sum\limits_{t=1}^{T}\mathbb{E}\left[
		V(\bm{x}^{t},\bm{\action}^{t};\bm{\reward}^{t})
		-\sum\limits_{n=1}^{N}\action_{n}^{t}y_{n}^{t}
		\right],
	\end{aligned}
\end{equation}
and introduce the second requirement: 
\begin{goal}\label{Goal: Principal_follow}
	To ensure the voluntary participation of the principal, the mechanism should satisfy $\mathtt{Pro}(T)\ge0$.
\end{goal}

\subsection{Fairness Requirement}\label{Subsection: Fairness}
The IOL framework relies on the selfish agents to play the arms for the principal.
In many networking applications (e.g., spectrum sharing), the arm-pulling actions are resource-consuming for the agents.
Besides the agent participation requirement, we follow the previous studies (e.g., \cite{kannan2017fairness,li2019combinatorial,gao2015providing}) and take into account the fairness of agent utilization.
We formulate the fairness requirement in  two steps:
\begin{itemize}
	\item First, we let $\thrFair_{n}\in[0,1]$ denote the maximal long-term utilization ratio of agent $n$.
	In practice, $\thrFair_{n}$ could be negotiated by the principal and agent $n$, or directly specified by the agent).
	Hence $\bm{\thrFair}=(\thrFair_{n}:\forall n\in\mathcal{N})$ is the public information in the IOL framework.
	The case $\thrFair_{n}=1$ means no fairness guarantee on agent $n$.
	
	\item Second, given the assignment $\bm{x}^{t}$ and the agents' decisions $\bm{a}^{t}$, we define the function $f_{n}(\bm{x}^{t},\bm{\action}^{t})$ in (\ref{Equ: f}) to indicate whether agent $n$ is actually utilized in slot $t$.
	Note that we have $f_{n}(\bm{x}^{t},\bm{\action}^{t})\in\{0,1\}$ for any  $\bm{x}^{t}\in\mathcal{X}$.
	\begin{equation}\label{Equ: f}
	f_{n}(\bm{x}^{t},\bm{\action}^{t})
	\triangleq
	\sum_{k=1}^{K}x_{n,k}^{t}\action_{n}^{t},\quad\forall n\in\mathcal{N}.
	\end{equation}
\end{itemize}

Based on the above discussions, we define the \textit{cumulative fairness violation} among the $N$ agents as follows:
\begin{equation}\label{Equ: condition_fair}
	\begin{aligned}
		\mathtt{Vio}(T)
		\triangleq
		\sum\limits_{n=1}^{N}\mathbb{E}
		\left(
		\sum\limits_{t=1}^{T}\Big[f_{n}(\bm{x}^{t},\bm{\action}^{t})-\thrFair_{n}\Big]\right)^+,
	\end{aligned}
\end{equation}
where $(\cdot)^+\triangleq\max(\cdot,0)$.
In general, the fairness violation $\mathtt{Vio}(T)$ is possibly increasing in $T$, while the sub-linear fairness violation (i.e., $\mathtt{Vio}(T)\sim o(T)$) implies the asymptotic satisfaction of the fairness requirement.
This is the third requirement in our mechanism design:
\begin{goal}\label{Goal: Fairness}
	The mechanism should ensure that the fairness violation is sub-linear in $T$, i.e., $\mathtt{Vio}(T)\sim o(T)$.	
\end{goal}


\subsection{Efficiency}\label{Subsection: Efficiency}
We evaluate the efficiency of an IOL mechanism based on the aggregated social welfare, which is the total profit (or payoff) of the principal and the agents.
In this problem, the social welfare is the difference between the principal's total reward $V(\bm{x}^{t},\bm{\action}^{t};\bm{\reward}^{t})$ defined in (\ref{Equ: valuation}) and the incurred agent cost $C(\bm{x}^{t},\bm{\action}^{t};\bm{\cost}^{t})$ defined as follows
\begin{equation}\label{Equ: Cost}
	C(\bm{x}^{t},\bm{\action}^{t};\bm{\cost}^{t})
	\triangleq
	\sum_{n=1}^{N}\sum_{k=1}^{K}
	\cost_{n,k}^{t}x_{n,k}^{t}\action_{n}^{t}.
\end{equation}
Accordingly, the social welfare in slot $t$ is given by
\begin{equation}\label{Equ: SW}
	\begin{aligned}
		S(\bm{x}^{t},\bm{\action}^{t};\bm{\reward}^{t},\bm{\cost}^{t})
		\triangleq  
		V(\bm{x}^{t},\bm{\action}^{t};\bm{\reward}^{t})
		-C(\bm{x}^{t},\bm{\action}^{t};\bm{\cost}^{t}).
	\end{aligned}
\end{equation}

We aim to maximize the cumulative social welfare in the long-term.
To understand the attainable social performance, we introduce two state-of-the-art benchmarks (i.e., $S^{*}$ and $S^{\dag}$) for the time-average social welfare in the following.
\begin{tcolorbox}
\begin{baseline}\label{Definition: benchmark}
	Suppose following two conditions hold:
	\begin{enumerate}
		\item[C1:] The average reward $\bm{\bar{\reward}}=(\bar{\reward}_{k}:\forall k\in\mathcal{K})$ is known.
		\item[C2:] Agents voluntarily participate (i.e., $\bm{a}^{t}=\bm{1}_{N}$) and disclose their private information in each slot.
	\end{enumerate}	 
	In the above idealistic setting, the maximal achievable social welfare is 
	\begin{equation}\label{Equ: benchmark}
		\begin{aligned}
			S^*\triangleq
			\max\limits_{\bm{x}(\cdot)}	
			&\  \mathbb{E}_{\bm{\cost}}\left[ \sum_{k=1}^{K}\sum_{n=1}^{N}\big(\bar{\reward}_{k}-\cost_{n,k}\big){x}_{n,k}(\bm{\cost}) \right] \\
			\textit{s.t. }	&\  \mathbb{E}_{\bm{\cost}}\left[f_{n}(\bm{{x}}(\bm{\cost}),\bm{1}_{N})\right]
			\le\thrFair_{n},\ \forall n\in\mathcal{N},
		\end{aligned}
	\end{equation}
	where the expectation is taken over the disclosed agent cost $\bm{\cost}=(\cost_{n,k}:\forall n\in\mathcal{N},k\in\mathcal{K})$.
	Moreover, $\bm{x}(\cdot)$ represents the policy (to be optimized).
	That is, the assignment $\bm{x}(\bm{\cost})\in\mathcal{X}$ will be adopted given the disclosed agent costs $\bm{\cost}$.
\end{baseline}
\end{tcolorbox}
Baseline \ref{Definition: benchmark} shows that achieving the social welfare benchmark $S^*$ requires two ideal conditions.
\textit{Condition C1} requires the prior knowledge on the uncertain environment.
Similar condition also appears in the benchmark definition of stochastic MAB \cite{bubeck2012regret}.
\textit{Condition C2} relaxes that fact that the agents are selfish.
This is the new condition for the IOL framework.
We measure the gap between $S^{*}$ and the social performance achieved by our mechanism based on the \textit{cumulative regret}: 
\begin{equation}\label{Equ: regret}
	\begin{aligned}
		\mathtt{Reg}(T)
		\triangleq
		\sum\limits_{t=1}^{T}\Big(
		S^{*}
		-
		\mathbb{E}\left[ S\left(\bm{x}^{t},\bm{\action}^{t};\bm{\reward}^{t},\bm{\cost}^{t}\right) \right]
		\Big),
	\end{aligned}
\end{equation}
where $\bm{x}^{t}$ and $\bm{a}^{t}$ denote the assignment and the agent decisions in slot $t$ under the IOL framework.
A sub-linear regret $\mathtt{Reg}(T)\sim o(T)$ implies that the proposed mechanism performs as well as the benchmark $S^*$ in the long-term.

The regret definition in (\ref{Equ: regret}) depends on the agents' strategic decisions and private information, thus it is a generalization of that adopted in stochastic MAB to the IOL framework.
It is already more challenging to achieve a sub-linear regret under the IOL framework.
Nevertheless, we will further introduce an even stronger social welfare benchmark in Baseline \ref{Definition: upper_bound}.
\begin{tcolorbox}
\begin{baseline}\label{Definition: upper_bound}
	Suppose that Conditions C1 and C2 in Baseline \ref{Definition: benchmark} hold, and the incurred agent cost is always the potential minimum value $\cost_{\min}$.
	In this idealistic setting, the maximal achievable social welfare is 
	\begin{equation}\label{Equ: upper_bound}
		\begin{aligned}
			S^{\dag}\triangleq
			\max_{\bm{p}}&\quad \sum_{k=1}^{K}\left(\bar{\reward}_{k}-\cost_{\min}\right)p_{k} \\
			\textit{s.t.}&\quad \sum_{k=1}^{K}p_{k}\le\Phi,\\
			&\quad p_{k}\in[0,1],\forall k\in\mathcal{K},
		\end{aligned}
	\end{equation}
	where $\bm{p}=(p_{k}:\forall k\in\mathcal{K})$ is the arm selection probability (to be optimized), and $\Phi=\sum_{n=1}^{N}\thrFair_{n}$ represents the average number of played arms per slot.
\end{baseline}
\end{tcolorbox}

Compared to Baseline \ref{Definition: benchmark}, Baseline \ref{Definition: upper_bound} relies on one more condition regarding the incurred agent cost.
Essentially, $S^{\dag}$ is the \textit{theoretical upper bound of the social performance}, since it is defined based on the potential minimal cost $\cost_{\min}$.
We will measure the performance gap between $S^{\dag}$ and our mechanism based on the following \textit{cumulative degradation}:
\begin{equation}\label{Equ: degradation}
	\begin{aligned}
		\mathtt{Deg}(T)
		\triangleq
		\sum\limits_{t=1}^{T}\Big(
		S^{\dag}
		-\mathbb{E}\left[ S\left(\bm{x}^{t},\bm{\action}^{t};\bm{\reward}^{t},\bm{\cost}^{t}\right) \right]
		\Big).
	\end{aligned}
\end{equation}

Note that one cannot change the cost $\bm{\cost}^{t}\in[\cost_{\min},1]^{N\times K}$ no matter how we design the learning and incentive scheme.
Hence there is an inevitable gap between the incurred agent cost (e.g., $\cost_{n,k}^{t}$) and the potential minimal cost $\cost_{\min}$. 
This makes it challenging to achieve a sub-linear degradation in general.
Surprisingly, in Section \ref{Subsection: Power of Crowd}, we will show that (under some mild condition) a sub-linear degradation is achievable when the agent crowd is large (i.e., $N$ is large).

The above discussions lead to the fourth requirement on the mechanism design problem for the IOL framework.
\begin{goal}\label{Goal: regret}
	The mechanism should achieve a sub-linear regret $\mathtt{Reg}(T)\sim o(T)$ and a sub-linear degradation $\mathtt{Deg}(T)\sim o(T)$.
\end{goal}

%


So far, we have formulated the mechanism design problem for the IOL framework based on Requirements \ref{Goal: agent_follow}$\sim$\ref{Goal: regret}.
A proper mechanism should manipulate the principal-agent interaction by providing incentives to agents and learning the unknown environment.
Next let us introduce the main result.

\section{Mechanism Design}\label{Section: mechanism design}
In this section, we propose a mechanism $\mathfrak{A}$ for the IOL framework.
For notation clarity, we let $(\bm{\hat{x}}^{t},\bm{\hat{y}}^{t})$ and $\bm{\hat{\action}}^{t}$ denote the proposal and agents' decisions in slot $t$ under our proposed mechanism $\mathfrak{A}$, respectively.
We first overview the basic idea in Section \ref{Subsection: Queuing-Based Control}.
We then introduce two major components of the mechanism $\mathfrak{A}$ in Section \ref{Subsection: Learning-Based Estimation} and Section \ref{Subsection: Auction-Based Elicitation}, respectively.
We summarize the entire mechanism in Section \ref{Subsection: Mechanism Description Complexity}.

\subsection{Overview of Mechanism $\mathfrak{A}$}
\label{Subsection: Queuing-Based Control}
In mechanism $\mathfrak{A}$, we construct a special Lagrangian function to facilitate the optimization of the proposal $(\bm{\hat{x}}^{t},\bm{\hat{y}}^{t})$ in each slot $t$.
The Lagrangian function in slot $t$ is
\begin{equation}\label{Equ: virtual SW}
	\begin{aligned}
		&\mathcal{L}\big(\bm{x},\bm{\QFair}^{t};\bm{\hat{\reward}}^{t},\bm{\hat{\cost}}^{t}\big)\triangleq \\
		&\qquad S(\bm{x},\bm{1}_{N};\bm{\hat{\reward}}^{t},\bm{\hat{\cost}}^{t}) -\sum_{n=1}^{N}\QFair_{n}^{t}\big[f_{n}(\bm{x},\bm{1}_{N})-\thrFair_{n}\big],
	\end{aligned}
\end{equation}
where $\QFair_{n}^{t}\ge0$ represents the Lagrangian multiplier associated with the fairness requirement on agent $n$ in slot $t$.
To better understand (\ref{Equ: virtual SW}), we further make two-fold elaborations:
\begin{itemize}
	\item In (\ref{Equ: virtual SW}), we presume that the agents are willing to follow the proposal $(\bm{\hat{x}}^{t},\bm{\hat{y}}^{t})$, thus substitute $\bm{\hat{\action}}^{t}=\bm{1}_{N}$.
	In Section \ref{Subsection: IOL_Performance}, we will show that this is indeed true under our proposed mechanism $\mathfrak{A}$.
	
	\item Recall that the reward vector $\bm{\reward}^{t}$ is prior unknown and the agent cost matrix $\bm{\cost}^{t}$ is agents' private information.
	Hence we define (\ref{Equ: virtual SW}) based on the estimated reward $\bm{\hat{\reward}}^{t}$ and the bidding cost $\bm{\hat{\cost}}^{t}$. We will elaborate how to generate $\bm{\hat{\reward}}^{t}$ and $\bm{\hat{\cost}}^{t}$ in Section \ref{Subsection: Learning-Based Estimation} and Section \ref{Subsection: Auction-Based Elicitation}, respectively.
\end{itemize}

Based on the Lagrangian function (\ref{Equ: virtual SW}), our proposed mechanism $\mathfrak{A}$ determines the assignment $\bm{\hat{x}}^{t}$ according to
\begin{equation}\label{Equ: assignment}
	\bm{\hat{x}}^{t}
	\triangleq
	\max\limits_{\bm{x}\in\mathcal{X}}\ \mathcal{L}\left(\bm{x},\bm{\QFair}^{t};\bm{\hat{\reward}}^{t},\bm{\hat{\cost}}^{t}\right),
\end{equation}
and the mechanism updates $\bm{\QFair}^{t+1}$ according to
\begin{equation}\label{Equ: Queue_fair}
	\bm{\QFair}^{t+1} 
	= 
	\Pi_{\mathbb{R}_{+}^{N}}\Big( \bm{\QFair}^{t} - \step\nabla_{\bm{\QFair}} \mathcal{L}\big(\bm{\hat{x}}^{t},\bm{\QFair}^{t};\bm{\hat{\reward}}^{t},\bm{\hat{\cost}}^{t}\big) \Big),
\end{equation}
where $\Pi_{\mathbb{R}_{+}^{N}}(\cdot)$ is the projection onto the non-negative orthant, and $\eta>0$ is the step-size (to be specified in Section \ref{Section: performance}).

Next we elaborate the rationale of generating $\bm{\hat{\reward}}^{t}$ and $\bm{\hat{\cost}}^{t}$ in Section \ref{Subsection: Learning-Based Estimation} and Section \ref{Subsection: Auction-Based Elicitation}, respectively.

\subsection{Estimated Reward}
\label{Subsection: Learning-Based Estimation}
Our proposed mechanism $\mathfrak{A}$ will estimate the reward primarily based on the Upper-Confidence-Bound (UCB) algorithm \cite{auer2002finite}.
Our goal in IOL mechanism design is not to improve this classic algorithm, but to design an appropriate incentive scheme that is compatible with this algorithm, so that they can properly manipulate the principal-agent interactions in the IOL framework together.
Next we introduce the details.

We let $H_{k}^{t}$ denote the number of observations on arm $k$ during the first $t$ slots, and let $\tilde{\reward}_k^{t}$ denote the empirical average reward of the $H_{k}^{t}$ observations.
At the beginning of slot $t$, mechanism $\mathfrak{A}$ generates $\bm{\hat{\reward}}^{t}=(\hat{\reward}_{k}^{t}:\forall k\in\mathcal{K})$ according to
\begin{equation}\label{Equ: UCB update rule}
	\begin{aligned}
		\hat{\reward}_k^{t}
		=
		\left\{
		\begin{aligned}
			&\min\left( \tilde{\reward}_k^{t-1} + \sqrt{ \frac{3\ln(t)}{2H_{k}^{t-1}} } , \ 1 \right),& H_{k}^{t-1}>0, \\
			& 1,& H_{k}^{t-1}=0.
		\end{aligned}
		\right.
	\end{aligned}
\end{equation}
When $H_{k}^{t-1}>0$, the empirical average reward $\tilde{\reward}_k^{t-1}$ represents the exploitation based on the past observations.
The term $[{\frac{3}{2}}\ln(t)/H_{k}^{t-1}]^{{1}/{2}}$ decreases in $H_{k}^{t-1}$ and represents the exploration out of the past observations.
When $H_{k}^{t-1}=0$, we set $\hat{\reward}_{k}^{t}$ as the potential maximal reward to prevent from overlooking a high-reward arm.

At the end of slot $t$, mechanism $\mathfrak{A}$ will update the counter $H_{k}^{t}$ and the empirical reward $\tilde{\reward}_{k}^{t}$.
Specifically, $\sum_{n=1}^{N}\hat{x}_{n,k}^{t}\hat{\action}_{n}^{t}\in\{0,1\}$ indicates whether arm $k$ is actually played by an agent in slot $t$.
Hence we have 
\begin{subequations}\label{Equ: val H}
	\begin{align}
		H_{k}^{t}&
		=H_{k}^{t-1}+\sum_{n=1}^{N}\hat{x}_{n,k}^{t}\hat{\action}_{n}^{t},\ \forall k\in\mathcal{K},\label{Equ: val h} \\
		\tilde{\reward}_k^{t}&
		= \left(\tilde{\reward}_k^{t-1} H_{k}^{t-1} + \reward_k^{t}\sum_{n=1}^{N}\hat{x}_{n,k}^{t}\hat{\action}_{n}^{t} \right)\bigg/H_{k}^{t},\ \forall k\in\mathcal{K}, \label{Equ: val emperical mean}
	\end{align}
\end{subequations}
where $\reward_{k}^{t}$ is the reward of playing arm $k$ in slot $t$.

\subsection{Bidding Cost}
\label{Subsection: Auction-Based Elicitation}
Our proposed mechanism $\mathfrak{A}$ manages to elicit the private agent cost through an auction scheme based on the Lagrangian function (\ref{Equ: virtual SW}).
Before introducing the auction procedure, for notation simplicity, we let $\mathcal{L}_{-n}(\bm{x},\bm{\QFair}^{t},\bm{\hat{\reward}}^{t},\bm{\hat{\cost}}^{t})$ denote the Lagrangian function at the absence of agent $n$, i.e.,
\begin{equation}\label{Equ: L -n}
	\begin{aligned}
		\mathcal{L}_{-n}(\bm{x},\bm{\QFair}^{t},\bm{\hat{\reward}}^{t},\bm{\hat{\cost}}^{t})
		\triangleq
		&\sum\limits_{k\in\mathcal{K}}\sum\limits_{i\ne n}\big(\hat{\reward}_{k}^{t}-\hat{\cost}_{i,k}^{t}\big){x}_{i,k}\\
		& -\sum\limits_{i\ne n}\QFair_{i}^{t}f_{i}(\bm{x},\bm{1}_{N}) +\sum\limits_{n\in\mathcal{N}}\QFair_{n}^{t}\thrFair_{n}.
	\end{aligned}
\end{equation}
Accordingly, we define $\mathcal{L}_{-n}^{\star}(\bm{\QFair}^{t},\bm{\hat{\reward}}^{t},\bm{\hat{\cost}}^{t})$ as the maximal value of  $\mathcal{L}_{-n}(\cdot,\bm{\QFair}^{t},\bm{\hat{\reward}}^{t},\bm{\hat{\cost}}^{t})$, i.e.,
\begin{equation}\label{Equ: L -n*}
	\mathcal{L}_{-n}^{\star}(\bm{\QFair}^{t},\bm{\hat{\reward}}^{t},\bm{\hat{\cost}}^{t})
	\triangleq
	\max\limits_{\bm{x}\in\mathcal{X}}\ \mathcal{L}_{-n}(\bm{x},\bm{\QFair}^{t},\bm{\hat{\reward}}^{t},\bm{\hat{\cost}}^{t}).
\end{equation}
Based on the above definitions, our proposed mechanism $\mathfrak{A}$ adopts the following \textit{Cost Revelation Auction}:
\begin{auction}[Cost Revelation Auction]\label{Auction: VCG}
	The auction procedure in mechanism $\mathfrak{A}$ consists of the following two steps:
	\begin{enumerate}
		\item Each agent $n\in\mathcal{N}$ submits the bids indicating his costs
		\begin{equation}
		\begin{aligned}
		\bm{\hat{\cost}}_{n}^{t}=(\hat{\cost}_{n,k}^{t}:\forall k\in\mathcal{K}).
		\end{aligned}
		\end{equation}
		\item The principle computes the payment $\bm{\hat{y}}^{t}=(\hat{y}_{n}^{t}:\forall n\in\mathcal{N})$ to agents according to
		\begin{subequations}\label{Equ: reward}
			\begin{align}
				\hat{y}_{n}^{t}
				\triangleq&
				\sum_{k=1}^{K}\left(\hat{\reward}_{k}^{t}-\QFair_{n}^{t}\right)\hat{x}_{n,k}^{t}\label{Equ: reward_1}
				\\
				&-\Big[\mathcal{L}_{-n}^{\star}(\bm{\QFair}^{t},\bm{\hat{\reward}}^{t},\bm{\hat{\cost}}^{t})-\mathcal{L}_{-n}(\bm{\hat{x}}^{t},\bm{\QFair}^{t},\bm{\hat{\reward}}^{t},\bm{\hat{\cost}}^{t})\Big],\label{Equ: reward_2}
			\end{align}
		\end{subequations}
		where $\hat{\bm{x}}^{t}$ is the adopted assignment in (\ref{Equ: assignment}).
	\end{enumerate}
\end{auction}

The payment rule (\ref{Equ: reward}) is the core of eliciting the private agent cost in mechanism $\mathfrak{A}$.
Note that the payment $\hat{y}_{n}^{t}$ to agent $n$ is independent of the bid $\bm{\hat{\cost}}_{n}^{t}$ of agent $n$.
Hence there is no incentive for each agent $n\in\mathcal{N}$ to misrepresent the costs in this slot.
Roughly speaking, the payment rule in (\ref{Equ: reward}) has two components with different roles:
\begin{itemize}
	\item The component (\ref{Equ: reward_1}) measures an agent's contribution to the principal under the fairness guarantee.
	According to (\ref{Equ: reward_1}), if agent $n$ plays arm $k$ (i.e., $\hat{x}_{n,k}^{t}=1$), then he will obtain the payment $\hat{\reward}_{n,k}^{t}-\QFair_{n}^{t}$.
	The estimated reward $\hat{\reward}_{n,k}^{t}$ represents agent $n$'s direct contribution to the principal under the assignment $\bm{\hat{x}}^{t}$.
	The Lagrangian multiplier $\QFair_{n}^{t}$ is an instantaneous monetary loss for ensuring the maximal long-term utilization ratio $\phi_{n}$.
	Overall, this component represents the contribution of an agent's presence and prevents the agent from being over-utilized.
	
	\item The component (\ref{Equ: reward_2}) is the loss inflicted on the others in the IOL framework by the presence of agent $n$.
	It follows from the basic VCG idea (e.g., \cite{vickrey1961counterspeculation,clarke1971multipart,groves1973incentives}) and measures the indirect effect of the agent.
	If an agent is irreplaceable in terms of being capable of pulling a high-reward arm that others cannot, then his presence will cause little effect on others, thus the loss in (\ref{Equ: reward_2}) is small.
\end{itemize}
As we will see in Section \ref{Section: Experiment Results}, the above payment rule makes it less costly for the principal to incentivize the individual when the crowd size $N$ increases.
This is an economic advantage originating from the power of crowd in the IOL framework.

\subsection{Mechanism Description}
\label{Subsection: Mechanism Description Complexity}
Algorithm \ref{Algorithm: A} summarizes our proposed mechanism $\mathfrak{A}$, which consists of the following phases in each slot $t$.
\begin{itemize}
	\item \textit{Lines \ref{A: auction}$\sim$\ref{A: leanring}:}
	After Auction \ref{Auction: VCG} is initialized, each agent submits the bidding cost $\bm{\hat{\cost}}_{n}^{t}$.
	Moreover, mechanism $\mathfrak{A}$ generates the current reward estimation $\bm{\hat{\reward}}^{t}$. 
	
	\item \textit{Lines \ref{A: decision}$\sim$\ref{A: decision_agent}:}
	Mechanism $\mathfrak{A}$ calculates the proposal $(\bm{\hat{x}}^{t},\bm{\hat{y}}^{t})$, and each agent decides whether to follow it.
	
	\item \textit{Lines \ref{A: val H}$\sim$\ref{A: final}:}
	Mechanism $\mathfrak{A}$ updates the counter $H_{k}^{t}$ and the empirical reward $\tilde{\reward}_{k}^{t}$ for each arm based on the new observation.
	We update the Lagrangian multiplier $\bm{\QFair}^{t+1}$ based on the assignment $\bm{\hat{x}}^{t}$ and agents' decisions $\bm{\hat{\action}}^{t}$.
\end{itemize}

\begin{algorithm}[t]
	\caption{IOL Mechanism $\mathfrak{A}$}\label{Algorithm: A} 
	\SetKwInOut{Input}{Input}
	\SetKwInOut{Output}{Output}  
	\Output{Proposal $(\bm{\hat{x}}^{t},\bm{\hat{y}}^{t})$ in each slot $t$.} 
	\textbf{Initial} $\step>0$, $\bm{\QFair}^{1}=\bm{0}_{N}$ and $H_{k}^{0}=0$ for any $k\in\mathcal{K}$.	\\
	\For {$t=1$ \KwTo $T$ } 
	{
		Initialize Auction \ref{Auction: VCG} and each agent $n\in\mathcal{N}$ bids $\bm{\hat{\cost}}_n^{t}$. \label{A: auction}
		
		
		Estimate reward $\bm{\hat{\reward}}^{t}$ according to (\ref{Equ: UCB update rule}).\label{A: leanring}
		
		Announce $(\bm{\hat{x}}^{t},\bm{\hat{y}}^{t})$ based on (\ref{Equ: assignment}) and (\ref{Equ: reward}).\label{A: decision}
		
		Each agent $n\in\mathcal{N}$ decides $\hat{\action}_{n}^{t}$. \label{A: decision_agent}
		
		%
		
		Update $H_{k}^{t}$ and $\tilde{\reward}_k^{t}$ for each arm $k\in\mathcal{K}$ based on (\ref{Equ: val H}).\label{A: val H}	
		
		Update $\bm{\QFair}^{t+1}$ based on (\ref{Equ: Queue_fair}). \label{A: final}	
	}
\end{algorithm}



So far, we have introduced mechanism $\mathfrak{A}$.
Next let us present its theoretical performance.

\section{Performance Analysis}\label{Section: performance}
We first proceed the general performance analysis for mechanism $\mathfrak{A}$ in Section \ref{Subsection: IOL_Performance}, and then investigate the impact of the crowd size $N$ in Section \ref{Subsection: Power of Crowd}.
The proof is given in our technical report \cite{TR}.

\subsection{Performance of IOL Mechanism $\mathfrak{A}$}
\label{Subsection: IOL_Performance}

\subsubsection{Instantaneous Performance}
The IOL framework relies on the selfish agents.
We first present how the selfish agents behave under our proposed mechanism $\mathfrak{A}$ in Theorem \ref{Theorem: Truthfulness}.
\begin{theorem}\label{Theorem: Truthfulness}
	In Auction \ref{Auction: VCG}, it is the dominant strategy for any  agent $n\in\mathcal{N}$ to truthfully bid his cost, i.e., $\bm{\hat{\cost}}_{n}^{t}=\bm{\cost}_{n}^{t}$.
	Furthermore, we have $\bm{\hat{\action}}^{t}=\bm{1}_{N}$ for any slot $t\in\mathcal{T}$.
\end{theorem}

Theorem \ref{Theorem: Truthfulness} has two-fold implications on the truthfulness and the voluntary participation of agents:
\begin{itemize}
	\item First, Theorem \ref{Theorem: Truthfulness} shows that truthfully bidding is no worse than any other bidding strategies for each agent, no matter how the others behave.
	Such a truthfulness notion corresponds to the dominant strategy implementation in each slot. 
	
	\item Second, Theorem \ref{Theorem: Truthfulness} implies that each agent achieves a non-negative payoff by following the proposal $(\bm{\hat{x}}^{t},\bm{\hat{y}}^{t})$ in mechanism $\mathfrak{A}$, thus we have $\hat{\bm{a}}^{t}=\bm{1}_{N}$.
	That is, mechanism $\mathfrak{A}$ ensures the voluntary participation.
\end{itemize}

The two aspects above enable our proposed mechanism $\mathfrak{A}$ to unconsciously enforce the desired assignment and achieve the other desired properties (to be introduced later).

\subsubsection{Long-term Performance}
We will present the long-term performance achieved by mechanism $\mathfrak{A}$ in Theorem \ref{Theorem: performance}.
For notation simplicity, we first define $\thrFair_{\min}$ as
\begin{equation}
\begin{aligned}
\thrFair_{\min}
\triangleq
\min_{n\in\mathcal{N}}\ \thrFair_{n},
\end{aligned}
\end{equation}
where $\thrFair_{n}$ is the maximal long-term utilization ratio of agent $n$.
Moreover, we define $\Xi(\delta)$ for any $\delta\in(0,\thrFair_{\min})$ as
\begin{equation}
	\Xi(\delta)
	\triangleq	
	\frac{3\sqrt{N}\Theta^2}{\thrFair_{\min}-\delta}\ln\left(\frac{2\Theta}{\thrFair_{\min}-\delta}\right) 
	+\frac{3\sqrt{N}\Theta}{2\delta},
\end{equation}
where $\Theta\triangleq\min(K+\Phi,N)$ is a constant, and $\Phi\triangleq\sum_{n=1}^{N}\thrFair_{n}$ represents the number of pulled arms per slot on average.
Theorem \ref{Theorem: performance} presents the long-term performance.
\begin{theorem}\label{Theorem: performance}
	With the step-size $\step=\frac{4K + 2\sqrt{6KT\Phi\ln T}}{T\Theta}$ in (\ref{Equ: Queue_fair}), mechanism $\mathfrak{A}$ in Algorithm \ref{Algorithm: A} attains
	\begin{subequations}\label{Equ: performance}
		\begin{align}
			&\mathtt{Reg}(T) 
			\le 6K+3\sqrt{6KT\Big[\Phi+\frac{\mathtt{Vio}(T)}{T}\Big]\ln T} ,\label{Equ: performance_reg}\\		
			&\mathtt{Vio}(T) 
			\le
			\Xi(\delta) 
			+\frac{\Theta^2}{4\delta}\sqrt{\frac{NT}{K\Phi}},\label{Equ: performance_vio}\\	
			&\mathtt{Pro}(T) + \frac{5K}{2}+2\sqrt{6KT\Big[\Phi+\frac{\mathtt{Vio}(T)}{T}\Big]\ln T}\ge0.\label{Equ: performance_profit}
		\end{align}
	\end{subequations}
\end{theorem}

The three inequalities in Theorem \ref{Theorem: performance} correspond to the social performance, the fairness violation, and the principal's participation, respectively.
We have two-fold elaborations:
\begin{itemize}
	\item First, (\ref{Equ: performance_reg}) and (\ref{Equ: performance_vio}) show that our proposed mechanism $\mathfrak{A}$ achieves a sub-linear regret $\mathtt{Reg}(T) \le O(\sqrt{T})$ and a sub-linear fairness violation $\mathtt{Vio}(T)\le O(\sqrt{T})$.
	That is, mechanism $\mathfrak{A}$ performs asymptotically  as well as the social welfare benchmark $S^{*}$ in Baseline \ref{Definition: benchmark}.

	\item Second, (\ref{Equ: performance_profit}) shows that the principal's  profit satisfies $\mathtt{Pro}(T)+O(\sqrt{T})\ge0$.
	This implies that our proposed mechanism $\mathfrak{A}$ enables the principal to obtain a non-negative profit in the long-term (i.e., $\lim_{T\rightarrow\infty}\mathtt{Pro}(T)\ge0$), which ensures the principal's voluntary participation.
\end{itemize}

Theorem \ref{Theorem: Truthfulness} and Theorem \ref{Theorem: performance} indicate that our proposed mechanism $\mathfrak{A}$ satisfies Requirements \ref{Goal: agent_follow}$\sim$\ref{Goal: regret} except the sub-linear degradation.
Recall that the degradation $\mathtt{Deg}(T)$ measures the social performance gap compared to the benchmark $S^{\dag}$, which is defined based on the potential minimal cost $\cost_{\min}$ in Baseline \ref{Definition: upper_bound}.
In practice, one cannot control the agents' costs, thus there is always an inevitable gap between the incurred agent cost and the potential minimal cost $\cost_{\min}$. 
Therefore, it is challenging to achieve a sub-linear degradation in the general setting.
Nevertheless, we will show that the sub-linear degradation is achievable in some specific scenario (when the agent crowd is large).
This is the \textbf{power of crowd} in the IOL framework.

\subsection{Power of Crowd in IOL Framework}\label{Subsection: Power of Crowd}
The IOL framework has a novel learning paradigm that relies on the crowd of agents.
Hence it is crucial to understand how the crowd size $N$ (i.e., the number of agents) affects the system performance.
To unveil the impact of $N$, we will narrow down our analysis to a statistically homogeneous IOL system $\mathcal{H}(\alpha,N)$, which is defined as follows:
\begin{definition}\label{Definition: System_N}
	The system $\mathcal{H}(\alpha,N)$ consists of $N$ selfish agents with the same maximal long-term utilization ratio, i.e., $\thrFair_{n}={\alpha}/{N}$ for any $n\in\mathcal{N}$.
	Moreover, the agents' costs $\{\bm{\cost}_{n}^{t}\in[\cost_{\min},1]^{K}:\forall n\in\mathcal{N},t\in\mathcal{T}\}$ are IID.\footnote{The agent cost realization and distribution are still the private information of the agents, thus are prior unknown to the principal.}
\end{definition}

Note that the system $\mathcal{H}(\alpha,N)$ focuses on $N$ agents who are statistically homogeneous in terms of the fairness requirement and the private cost distribution.
Moreover, there are $\sum_{n=1}^{N}\thrFair_{n}=\alpha$ arms played per slot on average in the above system $\mathcal{H}(\alpha,N)$.
Theorem \ref{Theorem: Deg} presents the main result of the cumulative degradation $\mathtt{Deg}(T)$.
\begin{theorem}\label{Theorem: Deg}
	In the system $\mathcal{H}(\alpha,N)$, there exists a finite constant $\epsilon>0$ such that mechanism $\mathfrak{A}$ attains the following cumulative degradation
	\begin{equation}\label{Equ: Deg_general}
		\begin{aligned}
			\mathtt{Deg}(T)
			\le
			\mathtt{Reg}(T)+ \frac{\alpha\epsilon T}{N+1},
		\end{aligned}
	\end{equation}
	where $\mathtt{Reg}(T)$ satisfies the inequality (\ref{Equ: performance_reg}) with $\Phi=\alpha$.
\end{theorem}

Theorem \ref{Theorem: Deg} presents a degradation upper bound, which is the sum between the regret $\mathtt{Reg}(T)$ and an extra term.
Specifically, the regret $\mathtt{Reg}(T)$ is sub-linear in $T$ according to Theorem \ref{Theorem: performance}, while $\frac{\alpha\epsilon T}{N+1}$ increases linearly in $T$.
Although the upper bound in (\ref{Equ: Deg_general}) is not sub-linear, we find that $\frac{\alpha\epsilon T}{N+1}$ decreases in the crowd size $N$.
This inspires us to investigate whether mechanism $\mathfrak{A}$ can achieve a sub-linear degradation when the agent crowd is large.
Corollary \ref{Corollary: Deg_More_Agents} provides the answer.
\begin{corollary}\label{Corollary: Deg_More_Agents}
	For any parameter $\beta\in(0,\frac{1}{3})$, in the system $\mathcal{H}(\alpha,N)$ satisfying $N=\lfloor \alpha^{\frac{1}{3}}T^{\beta} \rfloor$,
	mechanism $\mathfrak{A}$ attains the following performance
	\begin{subequations}\label{Equ: limiting}
		\begin{align}
			&\mathtt{Deg}(T)\le  O\left(\alpha^{\frac{2}{3}} T^{1-\beta}\right),\label{Equ: limiting_Deg}\\
			&\mathtt{Reg}(T)\le  O\left(\sqrt{\alpha KT\ln T}\right),\label{Equ: limiting_Reg}\\
			&\mathtt{Vio}(T)\le  O\left(\frac{K}{\alpha}\sqrt{KT^{1+3\beta}}\right),\label{Equ: limiting_Fair}\\
			&\mathtt{Pro}(T)+ O\left(\sqrt{\alpha KT\ln T}\right)\ge0.\label{Equ: limiting_Pro}
		\end{align}
	\end{subequations}
\end{corollary}

Corollary \ref{Corollary: Deg_More_Agents} presents the theoretical performance in the \textit{large-scale system}.
Let us elaborate this result in three steps:
\begin{itemize}
	\item First, the aforementioned large-scale system is formally defined by $N=\lfloor \alpha^{\frac{1}{3}}T^{\beta} \rfloor$.
	That is, given a total of $T$ operation periods, the crowd size $N$ should be sufficiently large.
	Such a large agent crowd enables our proposed mechanism $\mathfrak{A}$ to achieve a better social performance.
	
	\item Second, (\ref{Equ: limiting_Deg}) shows that mechanism $\mathfrak{A}$ achieves a \textit{sub-linear degradation} in the above large-scale system.
	This is the power of crowd in the IOL framework: a large crowd makes it possible to attain the theoretical upper bound $S^{\dag}$ in the long-term.
	This is the unique feature of the IOL framework.
	In Section \ref{Section: Experiment Results}, we will further demonstrate the economic impact of the crowd size $N$ based on the large-scale networking application.
	
	\item Third, (\ref{Equ: limiting_Reg})-(\ref{Equ: limiting_Pro}) present the other desired properties in the large-scale system.
	Specifically, the regret upper bound (\ref{Equ: limiting_Reg}) directly follows from (\ref{Equ: performance_reg}) when $\Phi=\alpha$. 
	Moreover, (\ref{Equ: limiting_Fair}) and (\ref{Equ: limiting_Pro}) validate the requirements on the agent fairness and principal participation, respectively.
\end{itemize}
To sum up, the power of crowd enables our proposed mechanism $\mathfrak{A}$ to satisfy Requirements \ref{Goal: agent_follow}$\sim$\ref{Goal: regret} in the large-scale system.

\begin{figure*}
	\centering
	\begin{minipage}{0.245\textwidth}
		\centering
		\setlength{\abovecaptionskip}{15pt}
		\setlength{\belowcaptionskip}{0pt}
		\includegraphics[width=0.99\linewidth]{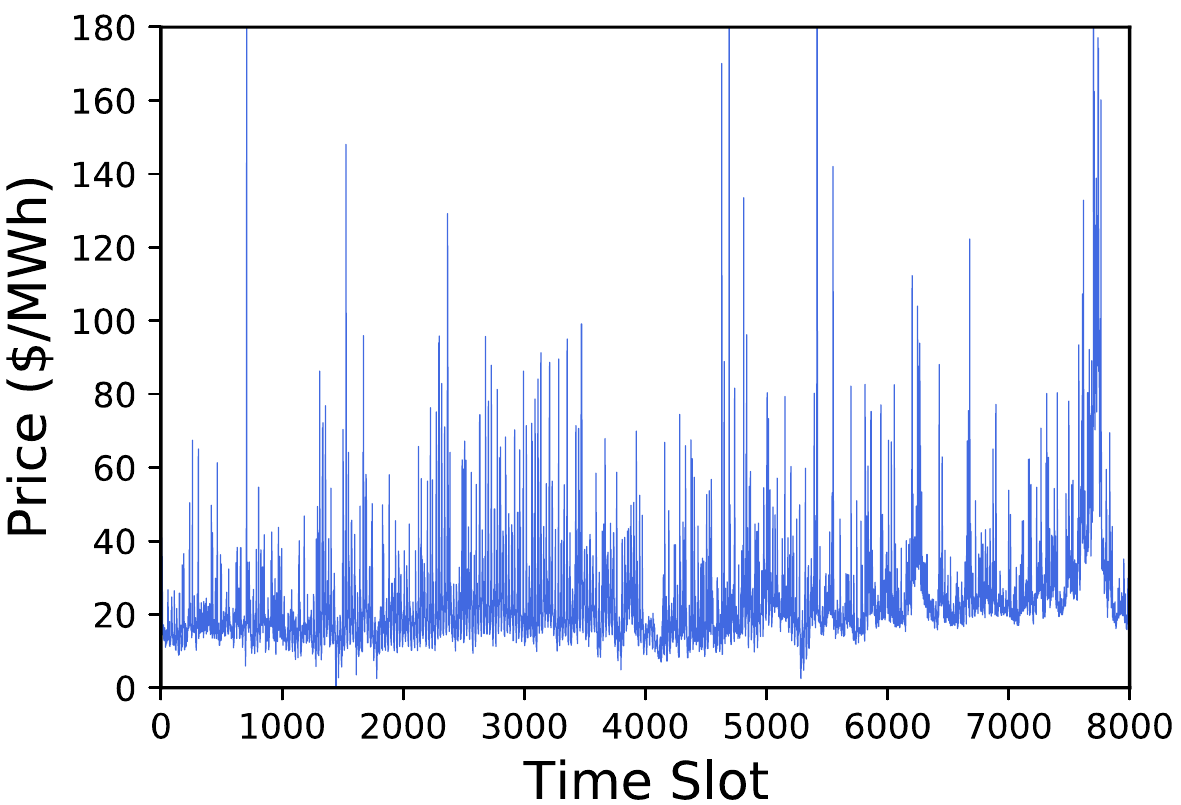}
		\caption{Electricity prices}
		\label{fig: price_IOL}
	\end{minipage}\ 
	\begin{minipage}{0.745\textwidth}
		\centering
		\setlength{\abovecaptionskip}{0pt}
		\setlength{\belowcaptionskip}{0pt}
		\subfigure[Social welfare]
		{\label{fig: SW}\includegraphics[height=0.228\linewidth]{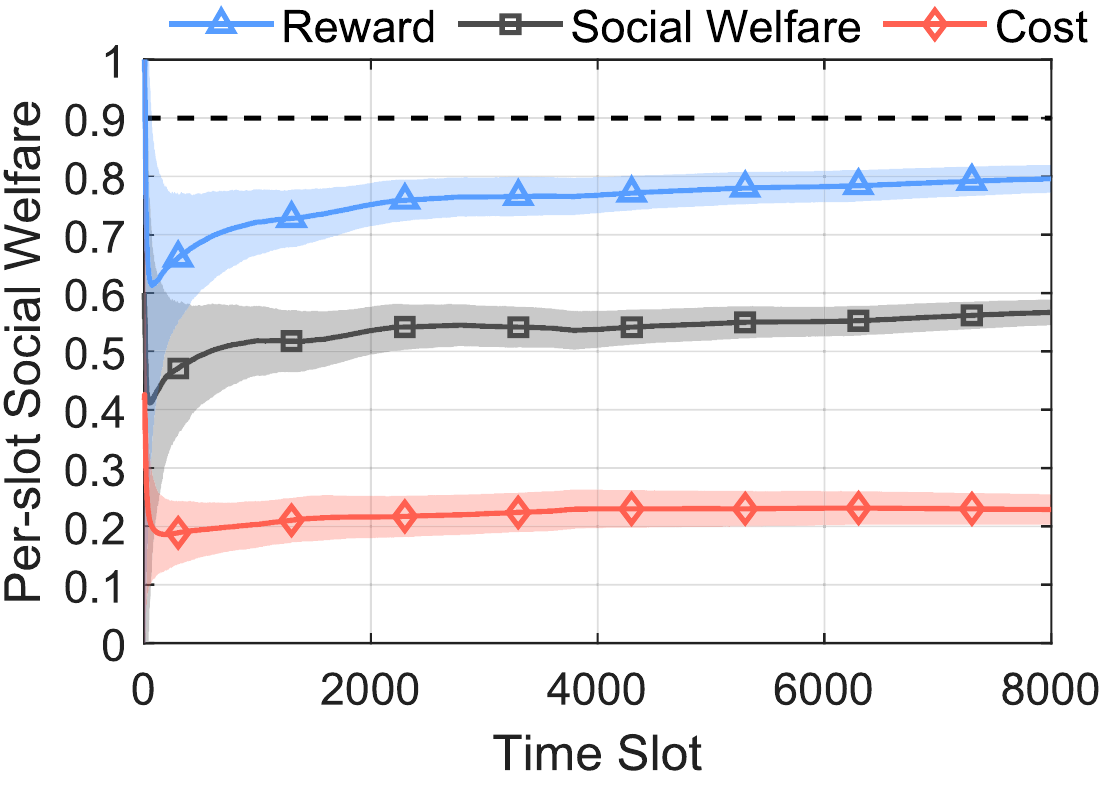}}
		\subfigure[Fairness violation]
		{\label{fig: Vio}\includegraphics[height=0.23\linewidth]{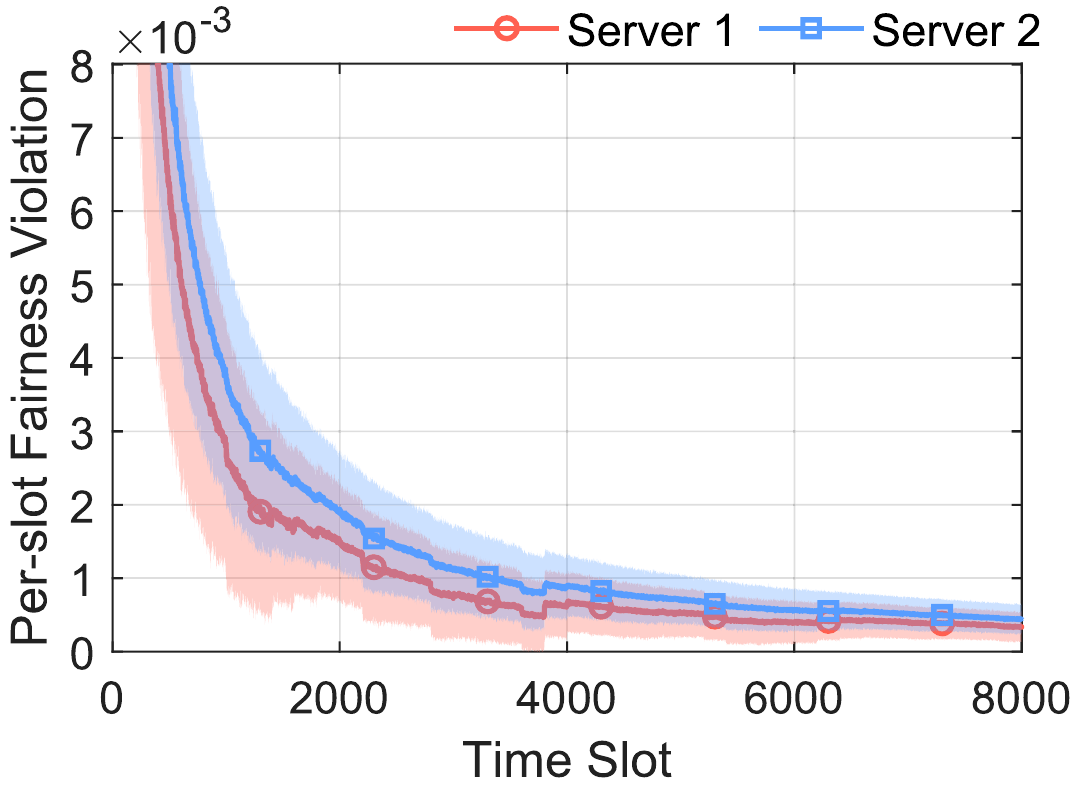}}
		\subfigure[Payoff and profit]
		{\label{fig: Payoff}\includegraphics[height=0.228\linewidth]{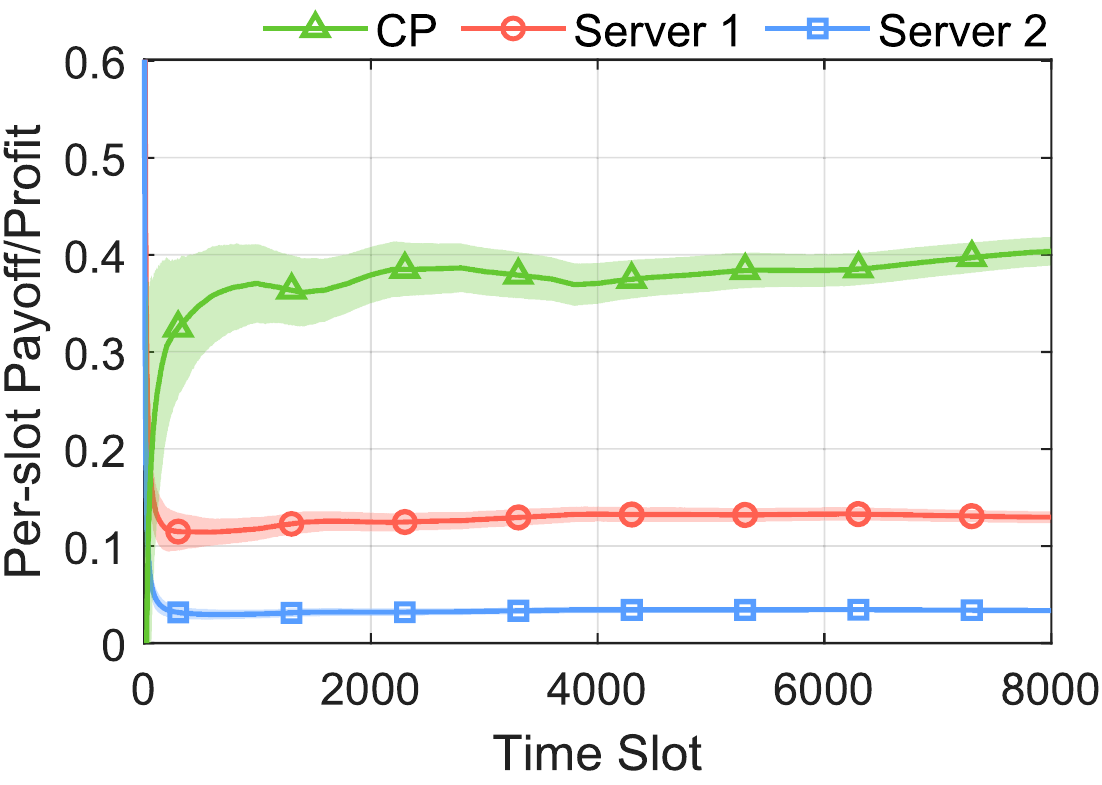}}
		\caption{Experimental results of the small-scale system}
		\label{fig: Small-scale}
	\end{minipage}
\end{figure*}


\section{Numerical Results}\label{Section: Experiment Results}
As an illustrative example, we will demonstrate how our proposed mechanism $\mathfrak{A}$ works in the edge computing network.
Let us start with characterizing the edge computing system via the IOL framework in Section \ref{Subsection: Edge Computing in IOL Framework}.

\subsection{Edge Computing in IOL Framework}
\label{Subsection: Edge Computing in IOL Framework}
Mobile Internet services (e.g., VR/AR) are becoming increasingly computation-intensive.
To improve the quality of experience (QoE), the Content Provider (CP) would incentivize the nearby edge servers to execute the computation tasks offloaded by the resource-limited mobile devices (that are using the CP's content services).
Next we elaborate this scenario based on the IOL framework.
\begin{itemize}
	\item \textbf{Principal:} The CP corresponds to the principal in the IOL framework.
	It aims to improve the QoE for mobile devices that are using its content services.
	
	\item \textbf{Arms:}	
	The mobile devices correspond to the $K$ arms in the IOL framework.
	They may have task offloading demands in certain time slot, which are unpredictable to the principal.
	
	\item \textbf{Agents:}
	The edge servers correspond to the $N$ selfish agents (with computing resource) in the IOL framework.
	They perceive a cost (i.e., energy expenditure) from executing computation tasks for mobile devices.
	Such a cost depends on the energy consuming rate and electricity prices, thus is the private information of edge servers.
\end{itemize}
Based on the above discussions, we model the reward $\reward_{k}^{t}\in\{0,1\}$ as the Bernoulli random variable, which indicates the task offloading of the mobile device $k\in\mathcal{K}$ in slot $t$.
That is, the CP perceives QoE improvement (i.e., reward) if one of the edge servers meets the task offloading demand of the mobile devices.
Furthermore, we model edge servers' costs as $\cost_{n,k}^{t}=\pi_{n}^{t}(\eta_{n}^{t}+l_{n,k}^{t})$, where $\pi_{n}^{t}$ represents the real-time electricity price.
Moreover, $\eta_{n}^{t}$ and $l_{n,k}^{t}$ measure edge server $n$'s energy consumption due to computation and communication (with mobile device $k$), respectively.

We first demonstrate how mechanism $\mathfrak{A}$ works based on a small-scale edge computing system in Section \ref{Subsection: Small-Scale Demonstration}.
We then evaluate the performance of the large-scale edge network in Section \ref{Subsection: Large-Scale Results}.

\subsection{Small-Scale Demonstration}
\label{Subsection: Small-Scale Demonstration}
We consider a small-scale edge network with $K=5$ mobile devices (i.e., arms) and $N=2$ edge servers (i.e., agents).
We set the maximal utilization ratio of the two servers as $\thrFair_{1}=0.7$ and $\thrFair_{2}=0.3$, respectively.
In this case, server 2 is expected to be less occupied than server 1, and on average $\thrFair_{1}+\thrFair_{2}=1$ server is ready to serve the mobile devices' task offloading.
As shown in Fig. \ref{fig: price_IOL}, our experiment follows the real-world electricity market price in US \cite{PJM}.
We randomly generate the energy consumption $\eta_{n}^{t}+l_{n,k}^{t}$ according to normal distribution truncated on the support $[0,0.1]$.
We use the average reward vector $\bm{\bar{\reward}}=\{0.1,0.3,0.5,0.7,0.9\}$ to capture the different offloading demands of the $K=5$ mobile devices.
Fig. \ref{fig: Small-scale} plots the per-slot average results of multiple simulation runs, where the shaded region represents the three-sigma range.

Fig. \ref{fig: SW} shows the per-slot average reward, cost, and social welfare achieved by mechanism $\mathfrak{A}$.
Overall, both the social welfare (i.e., the square curve) and the reward (i.e., the triangle curve) gradually increase.
However, there is still a significant gap between the reward (i.e., the triangle curve) and  $\bar{\reward}_{5}=0.9$ (i.e., the dash line).
Such a gap implies that incentivizing the selfish agents (i.e., the edge servers) prevents the principal (i.e., the CP) from choosing the optimal arm (i.e., the mobile device with the heaviest offloading demand).
Nevertheless, Section \ref{Subsection: Large-Scale Results} will show that our proposed mechanism $\mathfrak{A}$ can mitigate this drawback when the system scales.

Fig. \ref{fig: Vio} shows that the fairness violation is decreasing for the two edge servers.
Note that the cumulative violation for server 2 is a little larger than that for server 1, since server 2 corresponds to a smaller utilization ratio.
Hence the long-term payoff of server 2 is also smaller than server 1 as show in Fig. \ref{fig: Payoff}.
Furthermore, the triangle curve in Fig. \ref{fig: Payoff} shows that the CP's average profit is gradually increasing as mechanism $\mathfrak{A}$ learns about the average offloading demand.

\subsection{Large-Scale Evaluation}
\label{Subsection: Large-Scale Results}
\begin{figure}
	\centering
	\subfigure[]
	{\label{fig: PoC_SW}\includegraphics[height=0.35\linewidth]{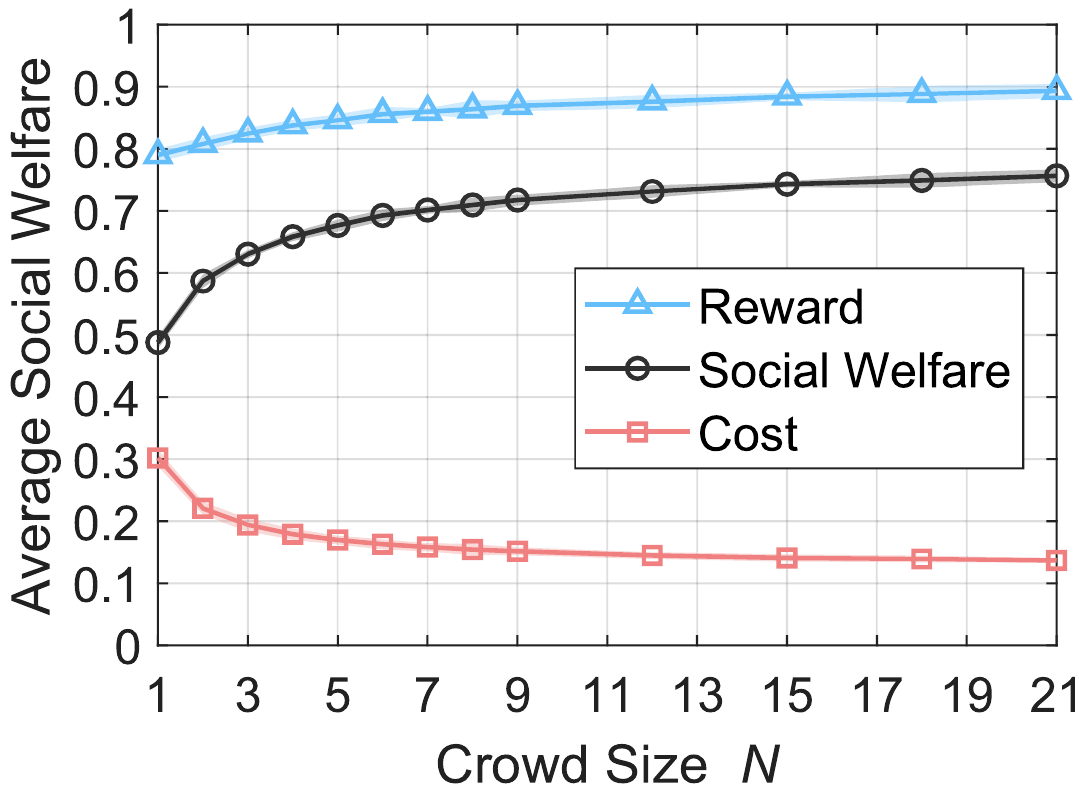}}\  \
	\subfigure[]
	{\label{fig: PoC_Payoff}\includegraphics[height=0.35\linewidth]{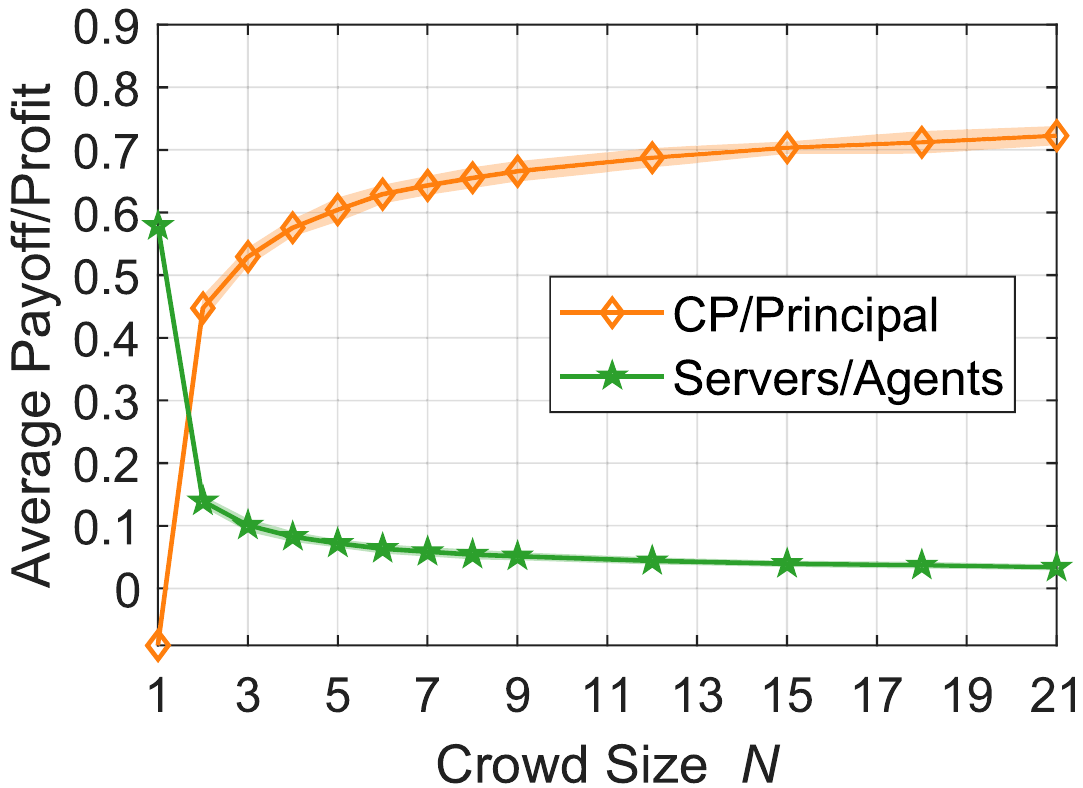}}
	\caption{Impact of the crowd size $N$.}
	\label{fig: PoC}
\end{figure}

Now we consider the large-scale edge network and investigate the impact of the crowd size $N$.
We set the parameters in Corollary \ref{Corollary: Deg_More_Agents} as $\alpha=1$ and $\beta=0.2$.
Fig. \ref{fig: PoC} plots the long-term average results for different crowd sizes.

Fig. \ref{fig: PoC_SW} shows that the total cost (i.e., the square curve) is decreasing in $N$, and the total reward (i.e., the triangle curve) is increasing in $N$ and converging to $\bar{\reward}_{5}=0.9$.
This is because that it would be more likely to match a low-cost edge server to the high-demand mobile device in a larger edge network.
Accordingly, a larger network would also improve the social performance as shown by the circle curve in Fig. \ref{fig: PoC_SW}.

Fig. \ref{fig: PoC_Payoff} shows that the CP profit (i.e., the diamond curve) is increasing in $N$, while the edge servers' average payoff (i.e., the star curve) is decreasing in $N$.
This is because that when there are more edge servers, each individual is becoming increasingly dispensable.
In this case, it incurs the CP less expense to incentivize the low-cost one according to the payment rule (\ref{Equ: reward}) in our proposed mechanism $\mathfrak{A}$.

\section{Conclusions and Future Works}\label{Section: Conclusion}
In this paper, we study the mechanism design problem for the incentivized online learning (IOL) framework.
The IOL framework exhibits a novel learning paradigm that relies on the selfish agents.
We characterize the strategic principal-agent interactions in this framework based on the stochastic MAB model.
We propose a mechanism that is incentive compatible and ensures the voluntary participation of agents and principal.
This mechanism performs asymptotically as well as the state-of-the-art baseline (that requires extra information) in terms of the social welfare and the fairness of agent utilization.
Our analysis also unveils the impact of the crowd size.
That is, in the large-scale system, our proposed mechanism can achieve the theoretical upper bound of the social welfare.

We view our study in this paper as the initial step towards understanding the IOL framework.
There are several valuable aspects for future investigations.
First, this paper focuses on the stationary environment.
It would be interesting to study the IOL mechanism design for the non-stationary case based on the adversarial MAB model.
Second, it is also interesting to study the revenue-maximizing mechanism design problem under the IOL framework.
Such a design objective would be more attractive to the principal.

\bibliographystyle{IEEEtran}
\bibliography{ref}

\end{document}